\documentclass[12pt,preprint]{aastex}
\usepackage{rotating}

\newcommand{\cmq}{cm{$^{-3}$}}

\newcommand{\msol}{M{$_{\odot}$}}

\newcommand{\kms}{km~s{$^{-1}$}}

\newcommand{\Hii}{H~{\sc ii}}
\newcommand{\Ha}{\mbox{H$\alpha$}}

\newcommand{\htwo}{H{$_2$}}

\newcommand{\CO}{{$^{12}$CO}}

\newcommand{\HCOP}{{HCO$^+$}}
\newcommand{\VLSR}{{V$_{LSR}$}}

\slugcomment{DRAFT: \today}
\shorttitle{ALMA observation of 114-426}
\shortauthors{Bally et al.}

\begin{document}

\title{ALMA Observations of the Largest  Proto-Planetary Disk
in the Orion Nebula, 114-426:  A CO  Silhouette }

\author{
John Bally\altaffilmark{1},
Rita K. Mann\altaffilmark{2},
Josh Eisner\altaffilmark{3},
Sean M. Andrews\altaffilmark{4},
James Di Francesco\altaffilmark{2,5}, \\
Meredith Hughes\altaffilmark{6}, 
Doug Johnstone\altaffilmark{2,5},
Brenda Matthews\altaffilmark{2,5},
Luca Ricci\altaffilmark{7}, \\
and 
Jonathan P. Williams\altaffilmark{8}}
	    
\affil{{$^1$}{\it{
      Department of Astrophysical and Planetary Sciences, \\
      University of Colorado, UCB 389, \\
      Boulder, CO 80309, USA}
      \email{john.bally@colorado.edu} }}
      
\affil{{$^2$}{\it{  
      National Research Council Canada, \\
      5071 West Saanich Road, \\
      Victoria, BC, V9E 2E7, Canada } } } 
        
\affil{{$^3$}{\it{
       Steward Observatory,  \\
       University of Arizona, \\
       933 North Cherry Ave., Tucson, AZ 85721, USA }}}
       
\affil{{$^4$}{\it{
     Harvard-Smithsonian Center for Astrophysics, \\
     60 Garden Street,  \\
     Cambridge, MA 02138, USA}}}
     
\affil{{$^5$}{\it{
      Department of Physics and Astronomy, \\
      University of Victoria, \\
      Victoria, BC, V8P 1A1, Canada }} }
      
 \affil{{$^{6}$}{\it{
      Van Vleck Observatory,  \\
      Astronomy Department, \\
      Wesleyan University, \\
      96 Foss Hill Drive,   Middletown, CT 06459, USA }} }
      
\affil{{$^{7}$}{\it{
      Department of Astronomy, California Institute of Technology, \\
      MC 249-17, Pasadena, CA 91125, USA   }} }
              
 \affil{{$^{8}$}{\it{
       Institute for Astronomy,  
       University of Hawaii, \\
       Honolulu, HI 96816, USA }}}

\begin{abstract}
We present ALMA observations of the largest protoplanetary disk in 
the Orion Nebula, 114-426.     Detectable 345 GHz   (856 $\mu$m) dust  
continuum is  produced only in the  350 AU  central region of 
the  $\sim$1000 AU diameter silhouette  seen against the bright  H$\alpha$ 
background in  HST images.    Assuming optically thin dust emission at 345 GHz,  
a gas-to-dust ratio of 100,  and a grain temperature of 20 K,  the disk gas-mass is 
estimated to be 3.1$\pm$0.6  Jupiter masses.   If most solids and ices 
have have  been incorporated into large grains, however, this value 
is a lower  limit.    The  disk is not detected  in dense-gas tracers  such as   
\HCOP\  J=4$-$3,   HCN J=4$-$3, or CS =7$-$6.    These 
results may indicate that the 114-426 disk is evolved and depleted in 
some light organic compounds found in molecular clouds.     
The CO J=3$-$2 line is  seen in absorption against the bright  50 to 80~K 
background  of the Orion A molecular cloud over the full  spatial extent 
and a little beyond the dust continuum emission.     The CO absorption  
reaches a depth of 27 K below the background CO emission at
 \VLSR $\,\approx$ 6.7 \kms\  $\sim\,$0.52\arcsec\   (210 AU) northeast and  
 12 K below the background CO emission at
\VLSR $\,\approx$ 9.7  \kms\  $\sim\,$0.34\arcsec\  (140 AU) southwest  of 
the suspected location of the central star, implying that the embedded star  
has a mass less than 1 \msol .     
\end{abstract}


\section{Introduction}

The Orion Nebula contains over a thousand low-mass young 
stars,  many of which  are surrounded by protoplanetary disks 
\citep{O'Dell2001}.   UV radiation from the massive Trapezium 
stars is photo-ablating these disks  to  produce 
several  hundred {\it proplyds} with comet-shaped, ionized skins 
seen in  \Ha\ and  visual-wavelength  forbidden emission-line 
images  \citep{OdellWen1994,Bally2000}.   In addition to these 
bright proplyds, the Nebula contains several dozen disks seen  
in silhouette against  the bright background of nebular light.  The 
absence of detectable  emission lines from some silhouettes
suggests that they are  either located far from the Orion Nebula's 
ionizing stars or completely outside the \Hii\ region in the foreground.

\begin{figure}
\epsscale{1.0}
\center{\includegraphics[width=0.9\textwidth,angle=0]
   {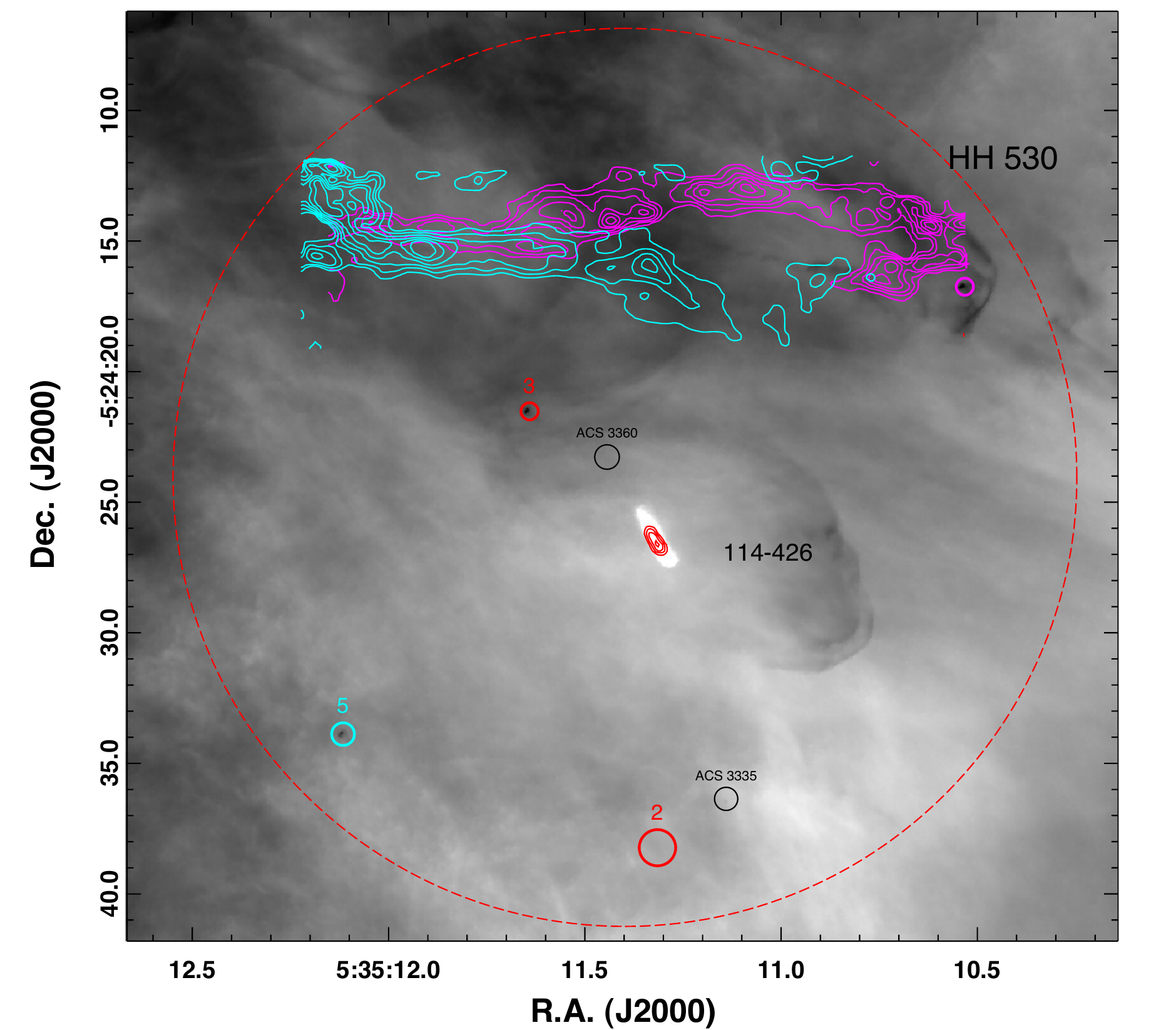}}
\caption{An HST \Ha\ image of the 114-426 field (Object 1 in Table~1).  
The large dashed circle 
shows the ALMA primary beam field of view at 350 GHz.  The numbered sources 
mark objects listed in Table~1 (except 114-426).    The red circles show two  objects (2, 3)
with  sub-mm continuum detections in addition to the 114-426 disk.  The cyan 
circle  marks the location  of object 5 (121-434 = OW94 = V2224 Ori),  
that was  not detected by ALMA.  
The  two black  circles mark additional stars detected in broad-band HST images
taken as part of the HST Treasury Program on the Orion Nebula Cluster in  Cycle 13, 
GO Program 10246, P. I.: M. Robberto \citep{Ricci2008,Robberto2013}.    
The CO outflow associated with HH 530 discussed in the text is shown in 
magenta and cyan contours.  Cyan  contours show blue-shifted emission 
in the radial velocity range \VLSR = 2 \kms\ to 5.9 \kms .  Magenta contours show  
red-shifted emission  in the radial velocity range \VLSR = 10.5 \kms\  to 14.3 \kms .
Contour levels are at 0.1, 0.2, 0.3, 0.4, 0.5, and 0.6 Jy beam$^{-1}$.  A magenta
circle an arc-second below the western end of the CO flow
marks the location of object 4 in the table, V2202 Ori 
%({\bf John, I think this should be numbered in the figure since that 
% is what you state at the top of the caption}).
% It is numbered in the Figure ... Unfortunately, the number is above 
% the circle where & blends with the magenta contours.
 }
\label{fig1}
\end{figure}

With a major axis diameter of  2.3\arcsec  , or 950 AU at a distance of  
414 pc \citep{Menten2007},  the 114-426  disk  is the largest  silhouette 
in  the Orion Nebula  \citep{McCaughrean1996}.      \citet{Throop2001}  
found  that the  translucent, northeastern rim has grey extinction to a   
wavelength of about 2 $\mu$m implying  that grains with  sizes of about 
1 $\mu$m or a little less dominate the extinction.  Reddening between 
1.87 $\mu$m and  4.05 $\mu$m, however,   suggests  that  most of the obscuration
is produced by grains smaller than 4 $\mu$m  \citep{Shuping2003}.  
\citet{Miotello2012}  argued that  despite being a silhouette,  UV-induced
photo-evaporation of the disk \citep{JohnstoneHollenbachBally98}
may be responsible for its  apparent warp,  the  translucent northeastern 
rim,   and the large-bow shaped  H$\alpha$ arc  6\arcsec\  to 10\arcsec\ 
west of the disk (see Figure \ref{fig1}).      

The nearly edge-on  disk  completely obscures  its central star  whose 
presence is indicated by a bipolar reflection nebula at visual and near-IR 
wavelengths.    \citet{McCaughrean1998}  used extinction of  background 
nebular light to estimate a lower-bound of   $2 \times 10^{-4}$ \msol\ for  
the mass  of the 114-426 disk.   They also used the  flux of scattered light  
in the reflection nebula to infer the intrinsic K-band magnitude of the central 
star, m$_K \sim$ 9.5,  and argued that it is likely to have a mass of $\sim$1.5 \msol .     
The extinction to the star was used to argue for a  disk mass 
larger  than $ 5 \times 10^{-4}$ \msol , higher than the lower
bound based on extinction of nebular light.  Multiple 
searches for  dust continuum emission at mm and sub-mm  wavelengths, however, 
have  failed to detect  emission  
\citep{Bally1998,Williams2005,Eisner2006,Eisner2008}.   
\citet{MannWilliams2009,MannWilliams2010} 
measured  the dust masses of several  dozen proplyds with the SMA 
interferometer,  but failed to detect 114-426, placing 
a limit of  $< 1.2 \times 10^{-2}$ \msol\  on the disk mass.
Finally, \citet{Mann2014} detected the dust continuum at 856~$\mu$m
with ALMA, finding a disk mass of about 3.4 Jupiter masses
($3.2 \times  10^{-3}$ \msol  ).

In this paper, we present the ALMA observations of the 114-426 disk   
in four spectral lines and re-analyze the dust continuum in more 
detail  than  \citet{Mann2014}.     Having a diameter of 
$\sim\,$950 AU,  114-426 is the largest candidate protoplanetary disk  
seen in silhouette against  the Orion Nebula. Nevertheless, 
the 856$\mu$m continuum is confined to the inner $\sim$ 350 AU region.     
Remarkably, carbon monoxide,   the only molecule detected in the 114-426 
disk,  is observed in absorption against the warm background CO emission 
from  the Orion~A molecular cloud.  The central star's mass is constrained by 
the disk rotation curve  to be less than 1 \msol .   The observations  show 
high-velocity  CO emission $\sim$10\arcsec\ north of 114-426,  possibly 
associated with  the HH 530 protostellar outflows in the Orion Nebula.

\section{Observations}

The data analyzed here were obtained with  ALMA 
during Cycle 0 (ALMA project 2011.0.00028.S) on 24 
October 2012 using twenty-two 12~m antennas, as 
part of a study of the Orion proplyds.   
\citet{Mann2014} present a detailed description of the observations 
and data reduction  procedures.   That paper also discusses  the disk 
dust masses measured using the  856 $\mu$m (350.0 GHz) dust 
continuum.    In this paper, we present  the molecular line data for 
the silhouette disk, 114-426.  Four transitions were observed in 
ALMA Band 7;  CO J=3$-$2, \HCOP\  J=4$-$3, HCN J=4$-$3, 
and  CS J=7$-$6.     Total on-source integration  times  were 1300 
seconds,  resulting in a continuum sensitivity of 
about 0.56 mJy per beam.  The synthesized beam full-width at 
half-maximum (FWHM) size is 0.51\arcsec\  ($\sim$211 AU) 
$\times$ 0.46\arcsec\ ($\sim$190 AU).   At this frequency, 1 Jy 
corresponds to a  brightness temperature of 42.8 K in the 
synthesized beam.    The primary-beam field of view is 18\arcsec\ 
and the maximum  recoverable angular structure is about 
5\arcsec\ ($\sim$2000 AU).     The correlator  was configured  
to observe four  1.875 GHz wide  bands using a  channel spacing 
of 488.28 kHz to yield  3840 channel spectra.    After Hanning smoothing, 
the spectral resolution was   $\Delta \nu$ 
=  976.56 kHz corresponding  to a velocity resolution 
$\Delta V$ = 0.84 \kms .   The spectral line sensitivity of these
observations is limited, presumably by over-resolved  fluctuations 
in the spatial and radial velocity structure of the spectral-line
emission from  the background molecular cloud,  to
about 0.025 to 0.1 Jy beam$^{-1}$ per Hanning-smoothed 
channel, or about  1 to 4 K per channel.    

The reduced data were re-gridded to  $X_{pix}$ = 0.11\arcsec\
per pixel.   Each pixel  value corresponds to the flux  that would be 
measured by the ALMA synthesized beam at the pixel center.     
The number of square  pixels with dimension $X_{pix}$  in a 
circular aperture which has the same  effective area on the sky 
as a Gaussian beam  with a (circular) full-width-half-maximum 
$FWHM$ = 0.485\arcsec\  is given by 
$N_{pix} =  [ \pi  / 4~ ln (2)] ( FWHM / X_{pix} )^2 \approx$ 22.03 
pixels\footnote{In terms of the FITS keywords $BMAJ$, $BMIN$, $CDELT1$,
   and $CDELT2$, the quantity 
   $(FWHM /  X_{pix} )^2$ is given by 
   $(BMAJ * BMIN) / (CDELT1 * CDELT2)$.
  } .  
Thus, the total flux in an aperture containing $n$ pixels is obtained by 
summing the pixel values in the aperture and dividing by $N_{pix}$.

\section{Results}
  
Figure \ref{fig1}  shows  the field-of-view of the ALMA 856 $\mu$m 
primary beam  (large dashed circle) superimposed on  a Hubble 
Space Telescope (HST) \Ha\  image  \citep{Bally2006}.    Red
contours show the dust  continuum  emission  from the 114-426 disk
on the visual wavelength silhouette.  High-velocity CO 
emission from a background protostellar outflow suspected 
to be associated with Herbig-Haro (HH) object HH 530  is  shown in
cyan  and magenta contours. 

Figure \ref{fig2} (top) shows a closeup of the disk in the  F775W  HST filter 
\citep{Ricci2008,Robberto2013}  with superimposed contours of 856 $\mu$m
continuum emission. The  sub-millimeter dust emission  has a major axis 
FWHM diameter of  1.11$\pm$0.05\arcsec , and a minor axis
FWHM diameter of 0.53$\pm$0.05\arcsec .    The beam 
de-convolved\footnote{The de-convolved major and minor axes were estimated
   by taking the square-root of the difference of the squares of the
   observed FWHM and beam FWHM.}
FWHM disk major  and minor axis diameters are 1.0\arcsec\ 
($\sim$ 414 AU) and less than 0.2\arcsec\ ($<$80 AU), 
implying that it is unresolved along its  minor axis.
Thus, the disk  outer radius at 856 $\mu$m is about   0.5\arcsec\  (200 AU),  
more than a  factor of two smaller than the  visual  wavelength  radius  
of 475 AU.      The peak  flux, 4.2 mJy beam$^{-1}$ is located at J2000 =
05:35:11.316, $-$5:24:26.62, coincident with the expected
position of the central star based on the symmetry of the bipolar reflection
nebula seen in the HST F775W filter image (Fig. \ref{fig2}). 
The area integrated flux in a  $\sim$ 0.74\arcsec\ by 1.55\arcsec\ diameter 
elliptical aperture aligned with the disk  major axis and centered on the emission 
is  6.0$\pm$0.6 mJy,    14\% smaller than the 7 mJy flux reported  by 
\citep{Mann2014}.      The  two flux measurements  differ slightly 
because of variations in the size and shape of the measurement aperture
combined with the presence of a low-amplitude negative bowl surrounding 
the continuum source.     The disk mass is computed from the measured 
flux density, $S_{\nu}$, the distance $D$  = 414 pc (Mann et al. 2014
used a distance of 400 pc), a grain emissivity
$\kappa$ = 0.034 ${\rm cm}^2~{\rm g}^{-1}$ (which assumes a gas-to-dust ratio
of 100), and a grain temperature $T_{dust}$ = 20 K, using the formula
$$
M_{disk}  = {{S_{\nu} D^2} \over {\kappa B_{\nu}(T_{dust}) }} 
\approx  5.9 \pm  1.1  \times 10^{30}~{\rm grams},
$$
where $B_{\nu}(T_{dust})$ is the Planck function.
Using these values, the disk mass is only 
3.1$\pm$0.6  times the mass of Jupiter . 

\citet{WilliamsBest2014} presented a parametric model for the estimation
of disk gas masses using isotopologues  of CO.   They found that 
for nine  disks they studied, the mass estimates based on the dust continuum 
flux using a standard gas-to-dust ratio of 100 were systematically higher than
the masses based on CO isotopologues.     Future ALMA observations of 
CO isotopologues should be obtained to test the validity of the assumptions
used in converting the sub-millimeter continuum flux into a disk gas mass.

Remarkably, the 114-426 disk is seen in {\it absorption} in the ALMA 
\CO\  J=3$-$2 images [Figures~2 (Bottom), 3, and 4 and Table~2].  The 
background CO emission from the  Orion A cloud is bright.     
In the Nobeyama 45 meter  15\arcsec\   telescope beam centered on 114-426,  
the \CO\  J=1$-$0  emission  has a  brightness  temperature above 50 K 
between \VLSR\ =   5 to 10 \kms\  with a peak value of  82 K at  \VLSR\ = 8.0 \kms\  
\citep{Shimajiri2011,Shimajiri2014}\footnote{
   The Nobeyama data was downloaded from the 
    NRO Star Formation Legacy Project website at 
    http://th.nao.ac.jp/MEMBER/nakamrfm/sflefaqcy/image.html\#orion}.
The IRAM 30-meter \CO\  J=2$-$1 emission \citep{Berne2014} shows  
peak temperatures  of 60$-$70 K  and the  APEX \CO\  J=3$-$2 map  
\citep{Peng2012} shows peak temperatures  of around 70 K 
behind 114-426.   
    \footnote{Since we do not have access to these data in digital 
     form,  our analysis relies on extrapolation of the lower-lying
     transitions.}     
These CO lines are optically 
thick  as indicated by the high intensities of  the $^{13}$CO and 
C$^{18}$O  lines at this location.   

The ALMA interferometer  filters-out structure on spatial scales larger 
than $\sim$5\arcsec\  in the configuration that was used for these 
observations.   Thus,  the mean total power in each frequency 
channel is zero.   Small-scale ($<$ 5\arcsec ) emission produces compact positive 
signals  surrounded  by extended (larger than 5\arcsec ) negative bowls.   
%Because ALMA filters out the low spatial-frequency components of the
%emission  in the field,   
Compact ($<$ 5\arcsec ) absorption features  with amplitudes larger than 
the small-scale intensity variations  in the background are retained in the 
maps as compact  regions with  negative values.    In the velocity range 
occupied by the strong background emission, the CO intensity fluctuations 
in the  ALMA   data cube have a 1~$\sigma$ rms  of about 0.1 Jy  or less
($\sim$4 K)  in apertures ranging in  diameter  from 0.5\arcsec\ to 5\arcsec ,  
less than 10\% of the single-dish surface brightness in the background 
line core against which  the CO absorption from 114-426 is seen,  
indicating that the background emission from the Orion A cloud is relatively 
smooth on these angular scales and does not contain intensity variations 
as large  as the absorption signal from 114-426.  

% ADD TEXT ON ABSORPTION BEING STILL VISIBLE

The northern portion  of the 114-426 disk (as traced by dust-continuum)  
has the strongest CO absorption with a depth of  $-$0.63 Jy beam$^{-1}$.   
The southern lobe of the dust continuum disk has a CO absorption 
depth of $-$0.28 Jy beam$^{-1}$  (Fig. 3).    
These flux deficits correspond to temperatures of 27 K  and 12 K below 
the background  CO emission.    Table~2 lists the location 
of the deepest  absorption as a function of radial velocity,  the absorption 
depth in Janskys and degrees Kelvin, the extrapolated single dish 
data  at the location of  114-426 interpolated to the radial-velocities
of the ALMA data, and the ratio of the absorption divided
by the estimated background CO temperature.    Figure~4 shows 
the CO data cube (grey-scale) with the outline of the visual silhouette
shown as a blue contour and the dust continuum emission shown in red
contours.   The  spatial extent of
the deepest,   blue-shifted CO absorption from the northeast part of the disk 
appears to be wider than the $\sim$ 0.5\arcsec\ 
synthesized ALMA beam and wider than the dust layer seen in the
HST images,  possibly indicating the presence of CO above and below the 
dust responsible for the visual-wavelength silhouette.  The deepest  
absorption in the northern portion of the disk  occurs at 6.7$\pm$0.15 \kms . 
The deepest absorption in the  southern portion of the disk occurs at 
9.6$\pm$0.15 \kms .    The two deepest absorption peaks differ 
in radial velocity by $\Delta$V = 2.9 $\pm$0.2 \kms\ and their 
centroids differ in position by  $\Delta$X = 0.86$\pm$0.1\arcsec  .   

\section{Discussion}

The observed CO absorption can be affected by several factors including 
the  optical depth, radial velocity, and temperature of the disk gas, 
beam dilution since the disk silhouette is thinner along 
its minor axis than the ALMA synthesized beam, and the brightness 
temperature distribution  of the background emission as a function 
of radial velocity.      

\begin{figure}
\epsscale{1.0}
\center{
 \includegraphics[width=0.74\textwidth,angle=0]
   {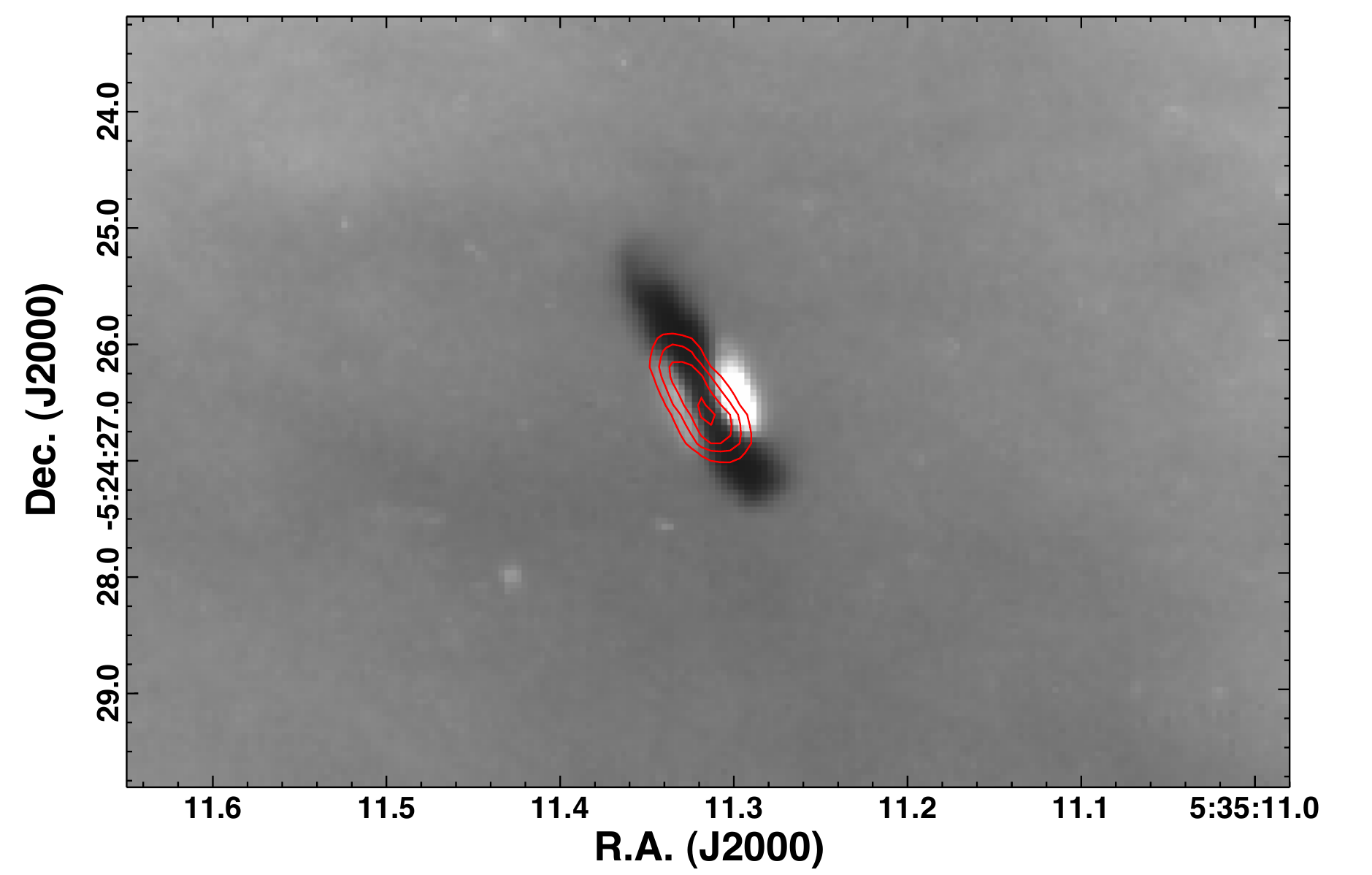}
 \includegraphics[width=0.75\textwidth,angle=0]
   {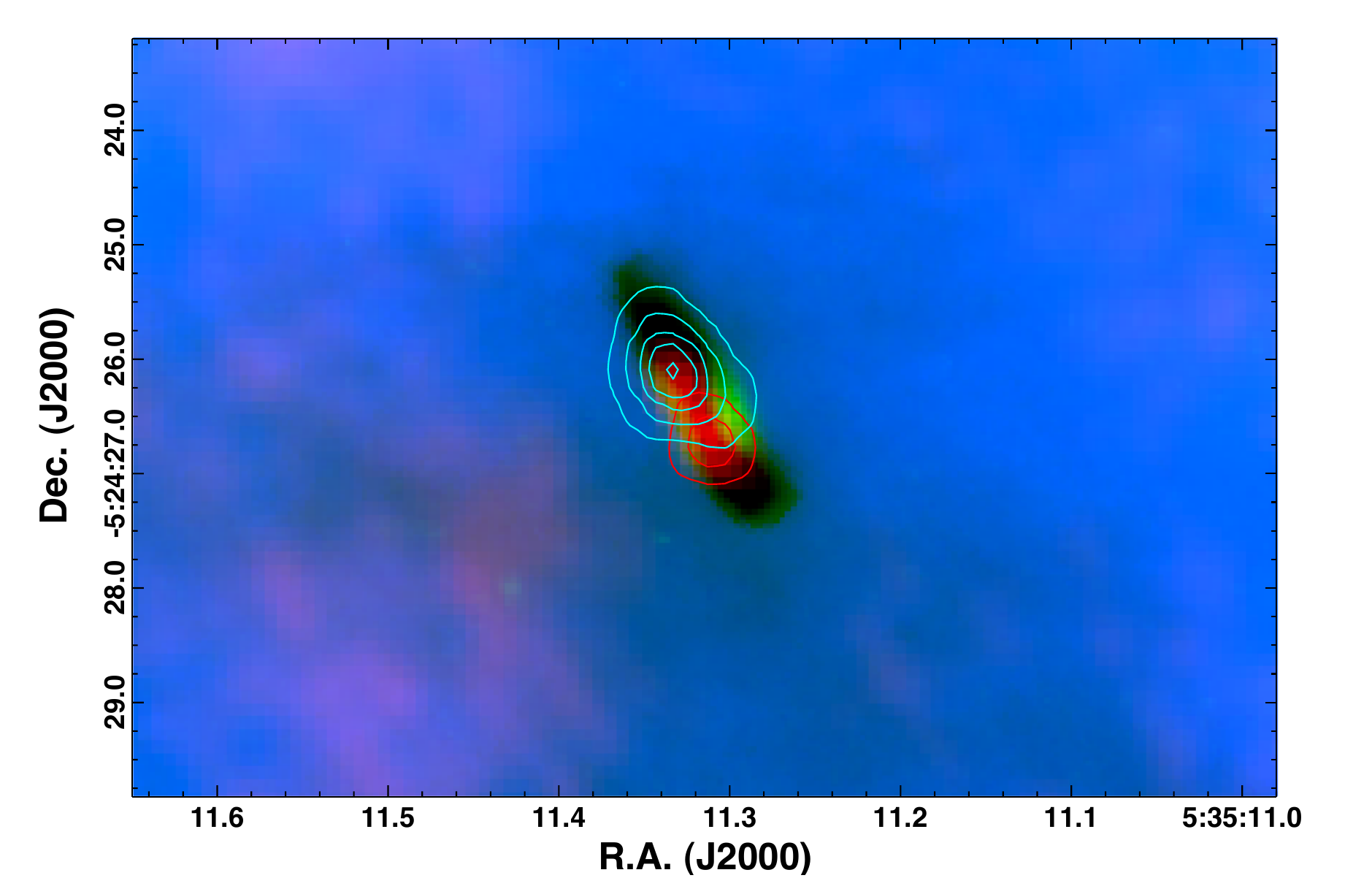}
  }
\caption{{\bf(Top):} 
A closeup showing an F775W image from the Hubble Space Telescope
\citep{Ricci2008} with contours of 856 $\mu$m dust continuum emission 
superimposed.  Contour levels are at 1, 2, 3, and 4 mJy beam$^{-1}$.
{\bf(Bottom):} 
A closeup of the 114-426 disk showing the 
H$\alpha$ image (blue) and the  F775W image (green)  
from the Hubble Space Telescope.  The dust continuum emission 
at  856 $\mu$m is shown (red) superimposed on the  HST image.
Contours of redshifted and blue-shifted CO J=3$-$2 absorption
at  \VLSR\ = 7.13 \kms\  (cyan contours) and at \VLSR\ = 9.24 \kms 
(red contours),  each in  0.42  \kms\  wide channels are
shown.
Contour levels are at  $-$0.2, $-$0.3, $-$0.4, $-$0.5, and $-$0.6, 
Jy beam$^{-1}$.
 }
\label{fig2}
\end{figure}

The projected 114-426 disk thickness varies from less than 
$z_d$ = 0.09\arcsec\  to 0.43\arcsec\  in the F775W  image (Fig.~\ref{fig2}).  
The smaller value, however,   corresponds to only the foreground portion
of the disk which occults the central reflection nebula.   
CO absorption against the background cloud samples the entire 
line-of-sight through the disk, 
including the portion behind the reflection nebula.   
The projected width  of the deepest CO absorption at 7.1 \kms\ 
(cyan contour in Fig. \ref{fig2} and the panel labeled 7.1 in Fig. \ref{fig4})
may be marginally wider than the 0.43\arcsec\  apparent  thickness of 
the silhouette.   As stated above, it is possible that optically thick CO extends 
above and below the region seen in silhouette in the HST images.

Emission and absorption from  the disk in the 
$r_B \sim\,$0.25\arcsec\  radius synthesized ALMA beam is diluted 
by a factor  $f_B \approx 2 r_B z_{CO} / \pi r^2_B$ =
$ 2 z_{CO} / \pi r_B$ $\sim$ 0.5 to 1.  
At the location of the strongest CO absorption in the northeast, the
thickness in the F775W  HST image is $z_d \approx$ 0.43\arcsec\  and 
$f_B \approx$ 1.  At the weaker southwest absorption, $z_d \approx$ 0.20\arcsec\  
and  $f_B \approx$ 0.5.   Correcting the southwest dip for beam dilution
makes its  absorption depth comparable to the northeast depth.
The inner part of the disk where the highest rotation speeds are expected
may be more beam diluted because the disk is expected to be even thinner.
Additionally, the highest velocity CO in  the center of the disk may
extend beyond the velocity range where the background emission
is brighter than the disk emission, in which case it would not be seen
in absorption and any emission signature may be too beam-diluted
to be seen.

The mean volume and column density of the 114-426  disk can be 
estimated from the dust-continuum derived mass, 
$M_{disk} = 5.9  \times 10^{30}$ grams,  and the projected size of
the sub-mm dust emission, $\sim$0.2\arcsec\ by 1\arcsec .
Assuming that the disk mass is distributed uniformly in a pill-shaped 
volume with these dimensions implies a mean density,
$n(H_2) > 10^7$ cm$^{-3}$ and column density, 
$N(H_2) > 4 \times 10^{22}$ cm$^{-2}$ .   The low-J transitions of
CO will be thermalized at the gas temperature, and unless it is depleted
by more than 3 orders of magnitude compared to molecular clouds,
it will be optically thick. 

The apparent surface brightness of the 856 $\mu$m continuum 
in the ALMA beam, $\approx$ 4 mJy beam$^{-1}$,  is also affected 
by beam dilution.   Using the beam  dilution factor  appropriate to the 
inner 350 AU portion of the disk,  $f_B \sim$ 0.5 to 1, implies an average 
surface brightness of  $S_{\nu} \sim$  8 $-$ 4  mJy if the disk were to fill the
beam.    Comparing this brightness to the Planck  function, $B_{\nu}(T)$  
for temperatures  ranging from  $T_{dust}$ = 20 $-$ 40 K from a solid-angle 
equal to the synthesized  ALMA  beam provides an estimate of the dust 
optical depth,  $\tau =  S_{\nu} / B_{\nu}(T)$.  The mean optical depth of 
the dust ranges from  $\tau$ = 0.016 to 0.032  for 20 K dust or 
$\tau$ = 0.006 to 0.013 for 40 K dust.

A surprising feature of the 114-426 dust continuum emission is the 
sharp drop in flux at a radius of about 200 AU from the suspected
location of the central star (Figs. \ref{fig1} and \ref{fig2}). 
The flux decreases by 3 mJy beam$^{-1}$ in a region comparable to or smaller
than the  synthesized beam diameter.  The outer parts of the disk beyond 
R $\sim$ 200 AU but interior to the edge at visual wavelengths 
(1.4\arcsec\ or 580  AU northeast of the suspected location of the central 
star and 0.8\arcsec\ or 330 AU to the southwest)  
do not exhibit any continuum emission 
above the noise.  Assuming that the total area of the outer region of
the disk is  0.4 square arc seconds based on a comparison between the
area of the silhouette in the HST images and the area where 856 $\mu$m
emission is detected, the 0.6 mJy/beam sensitivity limit implies that the column
density of  \htwo\ must be $N(H_2) < 8 \times 10^{21}$ \cmq.  This column corresponds 
to a visual-wavelength extinction of about 8 magnitudes, assuming normal ISM
dust.    The HST images show that the visual extinction through the outer 
disk  ranges from about 0.5 magnitudes in the translucent northeastern edge to
more than 3 magnitudes at R $\sim$ 200 AU.   Thus, for normal 
ISM dust, the visual extinction implies that the sub-millimeter  continuum 
ought to be detected at a level of  $\sim$0.2 mJy beam$^{-1}$ for a 0.5\arcsec\ beam 
diameter.
  
At the two velocities where the deepest CO absorption
is seen (6.9 \kms\ and 9.5 \kms ), the background CO line temperature is 
between 60 K and 70 K.     Assuming that the absorbing layer is optically 
thick and beam-filling (as it may be at the northwest portion of 114-426), 
an absorption depth of $-$0.6 Jy beam$^{-1}$ against a 
60 $-$ 70 K background corresponds to a brightness temperature 
of 33 $-$ 43 K.    If the CO absorption is beam-diluted by a factor $f_B$, 
the temperature will be  correspondingly lower and  likely a good estimate 
of the gas temperature where the CO optical depth reaches a value of 
order unity.

The CO absorption has a similar spatial extent as the  
dust continuum along the disk major axis.  
If the absorption minima trace the peak orbit speeds at their centroid
locations,  the mass of the central star can be estimated
since the 114-426 disk is within a few degrees of being 
edge-on.     As discussed above, the deepest absorptions occur 
at a projected separation  of   $\Delta X = 0.86 \pm 0.1$\arcsec ,   implying 
$R \approx 178 \pm 20$ AU  ($2.7 \pm 0.3 \times 10^{15}$ 
cm) at a distance of 414 pc.   The radial velocity difference between the 
north and south absorption dips, $\Delta V = 2.9 \pm 0.2$  \kms ,   
implies   $V_{orbit}(R)  = 1.45 \pm 0.1$ \kms .     Thus,  
assuming that the disk is in Keplerian rotaion,  the enclosed mass is  
$M_* = R V^2_{orbit}(R) / G \sim$ $0.43 \pm 0.11$ \msol  .   
This estimate may be a lower bound since  CO absorption would not 
trace gas outside the velocity range \VLSR\ = 5 $-$ 10 
\kms\  where the brightness temperature of the background CO emission 
drops below the brightness temperature of the CO in the disk.    The
absence of emission at larger radial velocities, however,  also sets 
an upper bound on the stellar mass.

We modeled the CO absorption from a geometrically thin, optically
thick Keplerian disk inclined by a few degrees, convolved with a Gaussian
to represent the synthesized ALMA beam using two free parameters:
an inner hole with radius $R_{in}$ and the central
star mass, $M_*$.  
%  Figure \ref{fig5} shows a $\chi ^2$ fit of a two-parameter model to the 
%  data as a function of the stellar mass and $R_{in}$.   
If the inner hole is small, a stellar mass between 0.4 msol\ and 0.7 \msol\
is favored.    A stellar mass of about 1 \msol\ and 
a thin annular disk with a large ($\sim$ 250 AU) hole, however,  
is also allowed.    A flaring disk with a small scale-height close the 
star  would produce results similar to a model disk with a large inner hole 
because of radially-dependent beam-dilution. 

Our  mass estimate is  lower than than 1.5 \msol\ 
estimated by \citet{McCaughrean1998} 
based on  extrapolation of the observed brightness of the reflection nebula. 
%m(K) = 14.5  magnitudes, to an m(K) = 9.5 magnitudes.   
More efficient
scattering by large, high-albedo grains, and significant forward
scattering would lower the estimated  intrinsic magnitude of the central star, and the 
resulting stellar mass estimate.  

\HCOP , HCN, and CS lines, in either emission or absorption,
are not detected from 114-426. Their brightness 
temperatures  may be either similar  to that of the background, 
or these species may be depleted.   Given the faint dust continuum
emission, and the low relative abundances of these species compared to CO
in molecular clouds \citep{BerginTafalla2007}, their absence in 114-426
may not be that surprising.     

\subsection{Is 114-426 an Evolved Proto-planetary Disk?}

The small estimated mass of the 114-426 disk is in stark contrast to its large 
size in visual wavelength images.    Visual extinction is dominated by small 
grains while the sub-millimeter emission  is more sensitive to large  particles.  
Multi-wavelength observations of circumstellar disks indicate that the grain size 
decreases with increasing distance from the central star.    These observations
suggest that the outermost regions of young circumstellar disks might be populated 
by dust similar to the ISM \citep{Testi2014}. 
  
Photometry of the translucent northeast outer edge of the 114-426 disk 
has shown that the extinction is grey, implying that the mean particle size 
responsible for the attenuation of background light is large compared to 
interstellar grains in the outer parts of the disk at a projected radius of 400 AU 
to 500 AU from its central star \citep{Throop2001}.   Reddening 
between  2 $\mu$m and 4 $\mu$m \citep{Shuping2003}, however,  indicates 
that the grains must be smaller than a few micro-meters.   These observations 
suggest that the 114-426 is an evolved disk in which solids and ices have been 
incorporated into large particles.     These observations, and comparison to
data on young disks \citep{Testi2014}, suggests that 114-426 is an evolved
disk.

Grain growth rates are expected to increase with decreasing distance from 
the central star.   Thus, it is possible that in the inner portions of the 114-426
disk, particles have grown to sizes larger than a millimeter, in which case the
sub-millimeter  emissivity per unit mass of solids (and gas) would decrease.
Thus, the assumptions used above for converting the observed 
continuum flux into dust and gas mass may not apply.  In this case, our estimated 
disk mass would be a lower bound.   If the mean particle size responsible
for the continuum emission is larger than the wavelength, however, then the shape 
of  the dust continuum spectrum should approach the Rayleigh-Jeans slope of
$-$2.  Future sub-arc-second angular resolution, multi-frequency observations
with ALMA can directly measure the sub-millimeter  continuum spectral index to
determine if indeed grains are larger than the observing wavelength.   Given that
the 114-426 disk lacks evidence for being surrounded by a photo-ionized skin, 
spectral index measurements can be extended into the centimeter-wavelength
domain using the JVLA.  Such measurements may in the future provide direct
evidence for the sequestration of solids and ices into large bodies, a process 
expected to be the first phase of planet formation.

It is possible that the absence of dense gas tracers in the 114-426 disk is also
evidence for chemical evolution.  In the low-temperature conditions encountered in
the outer Orion Nebula, many common tracers of molecular gas may have frozen 
out of the gas phase and locked into icy grain mantles.  UV and X-ray photolysis
of light organic compounds in an ice matrix may lead to their conversion into
heavy organic material.  Thus, the absence of measurable HCN, CS, and \HCOP\
lines may also be an indication that the 114-426 disk is highly evolved.     Sensitive
searches for other compounds in the sub-millimeter spectrum of 114-426 with ALMA
should test this hypothesis.  

\citet{Miotello2012} proposed that 114-426 is being photo-ablated by the non-ionizing
FUV radiation field in the Orion Nebula and suggested a mass-loss rate 
$\dot M_{disk} \sim (3 -5) \times 10^{-7}$ \msol\  yr$^{-1}$.  For the disk mass 
estimated above, $5.9 \times 10^{30}$ grams, the disk lifetime is 
$\tau \sim M_{disk}  / \dot M_{disk}  \sim (1.0 - 1.6) \times 10^4$ years.

\subsection{Comparison to ALMA Observations of Other Silhouette Disks in Orion}

Two other  giant disks in Orion,  216-0939 near OMC2, and 
253-1536 in M43  show strong dust continuum \citep{Mann2014}
and bright line-emission in the dense gas tracers.   CO is seen both in
emission and absorption, but the latter signal is  less prominent  than in 114-426,  
probably because the background CO emission is dimmer, and possibly
because  foreground CO is more likely to be present along the line of sight. 
Analysis of the displacements and Doppler shifts of the absorption signals give
stellar mass estimates comparable to those derived from the emission line 
rotation curves, namely 3.5 \msol\ for  253-1536 \citep{Williams2014} and 1.1 
\msol\ for 216-0939 (in preparation).    These objects 
have central star masses around 1 \msol\  to 3 \msol\ and are surrounded by massive
disks  exhibiting strong sub-mimmimeter dust continuum and bright line emission in 
HCN, CS, and HCO$^+$.    The dim dust continuum and lack of detection of
the dense gas tracers in the 114-426 silhouette disk is consistent with the 
low mass estimated for its central star and surrounding disk 
\citep{Andrews2013}.

\subsection{Other Objects in the 114-426 ALMA field}

Figure 1 shows that the northern portion of the 114-426 ALMA field 
contains  bright CO emission associated with the Herbig-Haro
object HH~530 \citep{Bally2000}.   The H$\alpha$ image  
shows a jet-like feature  east of the star V2202 Ori 
(source 4 in Table~1 and the figures; source 106-417 in O'Dell and Wen 1994).  
Many components of HH~530, however,  exhibit supersonic motions towards
the southwest with velocities ranging from 25 \kms\  to 70 \kms\ \citep{ODellDoi2003},
indicating that these shocks trace a background outflow from a source embedded
in the vicinity of OMC1-South, northeast of the 114-426 field.   
Thus, V2202 Ori  is not related to this flow.
The CO  emission in the vicinity of HH~530 extends from \VLSR\ = +2  \kms\  to 
over +20 \kms\  and may be  associated with HH~530.  It
may trace a shell swept-up by this externally irradiated HH object.   
Given that both red and  blueshifted components of the flow are seen in the same lobe
(with respect to the \VLSR\ $\sim$ 9 \kms\  radial velocity of the background 
CO emission in this portion of Orion),  the outflow  must move close to the 
plane of the sky.    The presence of molecules in HH~530 shows that the H$\alpha$
emission seen in the HST images must trace the ionized skin of a molecular outflow,
indicating that this flow is only partially photo-ionized by the UV radiation field 
in the Orion Nebula.

The prominent bow-shaped \Ha\ structure located about 
6\arcsec\ to 10\arcsec\ west of 114-426 is not associated with any CO 
features.  This bow, however,  also exhibits a large proper motion of about 
40 \kms\ toward the southwest as shown by the multi-epoch HST images of
\citet{ODellDoi2003}.     It is unclear if this  feature is associated with an 
outflow from 114-426 or an other unrelated background  flow similar to HH~530 
but without a CO counterpart.

\section{Conclusions}

The largest disk seen in silhouette against the visual-wavelength emission
of the Orion Nebula, 114-426, is detected in dust continuum emission, and
in absorption in the 345 GHz J=3$-$2 CO transition.    The CO absorption against
the warm background emission from the Orion A cloud has the same spatial extent
and morphology as the dust continuum emission.   The disk
in these tracers  is a factor of two smaller than in the visual wavelength HST images, however. 
A  model consisting of a Keplerian rotation curve in a geometrically-thin, edge-on disk
convolved with a 0.5\arcsec\  Gaussian beam-profile indicates that  $M_*$ is less
than 1 \msol\  with a most likely value around 0.4 \msol.     
The disk is not detected in HCN, CS, or HCO$^+$ lines. 

Assuming a gas-to-dust ratio of 100, and a a dust temperature of 20 K, 
the observed sub-mm dust continuum implies a disk mass of  3.1$\pm$0.6  
times the mass of Jupiter  ($5.9 \pm  1.1  \times 10^{30}$ grams).   
Comparison with ALMA observations of two other large disks in the Orion 
region provides support for a low-stellar mass  and low-disk mass.

\acknowledgments{
This research of JB was in part supported by  National Science 
Foundation (NSF) grant  AST-1009847. DJ acknowledges support from a 
Natural Sciences and Engineering Research Council (NSERC) Discovery Grant. 
This paper uses ALMA data obtained with program  ADS/JAO.ALMA\#2011.0.00028.S. 
ALMA is a partnership of the European Southern Observatory (ESO) 
representing member states, Associated Universities Incorporated (AUI) and the National
radio Astronomy Observatories (NRAO) for the National Science Foundation (NSF) in the USA,  
NINS  in Japan, NRC in Canada, 
and NSC and ASIAA in Taiwan, in cooperation with the Republic of Chile.  The Joint ALMA
Observatory (JAO) is operated by ESO (Europe) , AUI/NRAO  (USA), and NAOJ (Japan).
The National Radio Astronomy Observatory is a facility of the National Science Foundation 
operated under cooperative agreement by Associated Universities, Inc.
 }

\clearpage

\bibliography{ms}

\clearpage

\begin{figure}
\epsscale{1.0}
\center{\includegraphics[width=1.0\textwidth,angle=0]
   {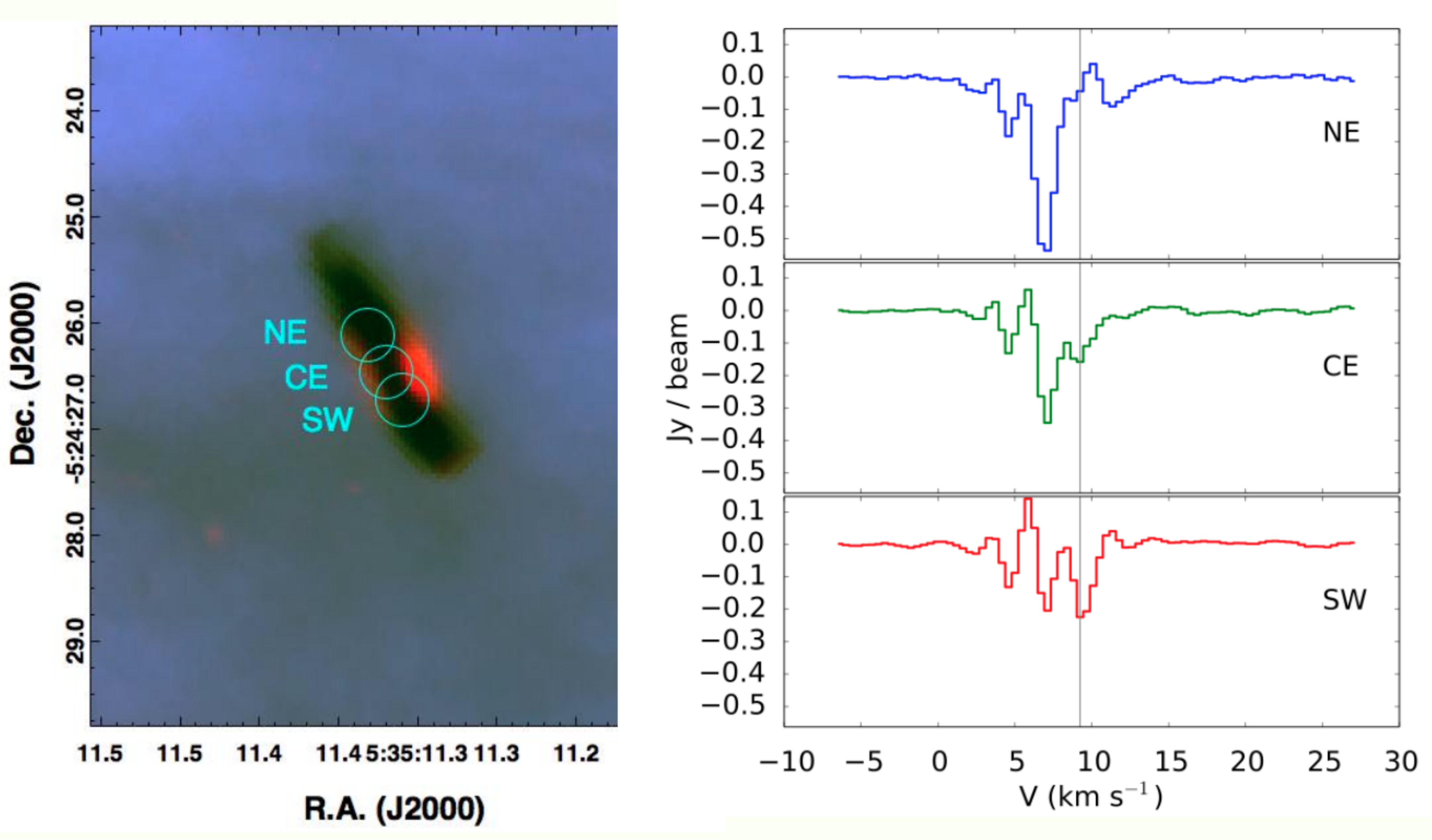}}
\caption{
{\bf(Left):} The H$\alpha$  (blue and green) and the F775W  (red) HST images
combined with the ALMA synthesized beams superimposed.
{\bf(Right):}   ALMA J = 3-2 CO spectra showing the absorption at the three positions
indicated in the left panel where CE refers to the central position.  The vertical
line marks the approximate central velocity of the background emission,  
V$_{LSR}$ = 9.2 km~s$^{-1}$.
 }
\label{fig3}
\end{figure}

\begin{figure}
\epsscale{1.0}
\center{\includegraphics[width=1.0\textwidth,angle=0]
   {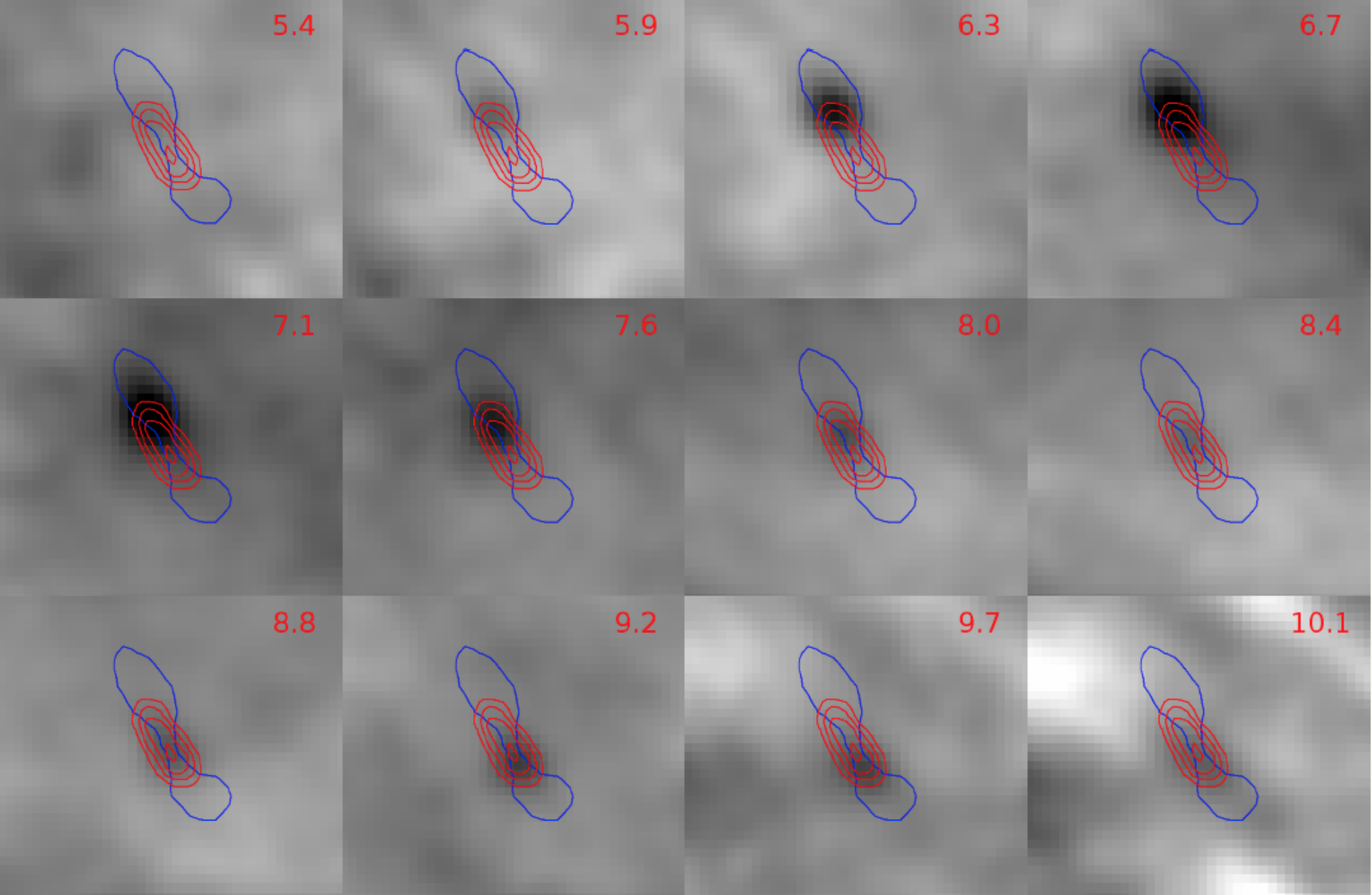}}
\caption{ 
A grid of images showing the 114-426 disk in the CO J = 3-2 line (grey scale).  The 
data values shown range over  $-$1.0 $-$ +1.0 Jy beam$^{-1}$.  
The LSR radial velocity of the center of each 
channel is indicated in the upper right in km~s$^{-1}$.  The blue contour shows the
outline of the disk in the HST F775W filter.  The red contours show the 856 $\mu$m
dust continuum emission with the same levels as in the top panel in Figure 2.
 }
\label{fig4}
\end{figure}

% \begin{figure}
% \epsscale{1.0}
% \center{\includegraphics[width=0.9\textwidth,angle=0]
%    {ms_f5_chi2_smooth.pdf}}
%  \caption{A contour plot of the $\chi ^2$ fit to the data showing the stellar mass
%  as a function of the inner-radius of the circumstellar disk.  Black is the 1 $\sigma$, 
%  dark grey is the 2 $\sigma$, and light grey is the  3 $\sigma$ allowed region.  
 %  }
%  \label{fig5}
%  \end{figure}

\clearpage

%xxxxxxxxd114-426  2 October 2014 xxxxx

\begin{deluxetable}{l c c r l}
 %\scriptsize{10pt}
 % \tabletypesize{\scriptsize}
 %\setlength{\tabcolsep}{0.02in}
  %\tablewidth{0pt}
  
\centering

\tablecaption{ 
      Sources in the d114-426 Field 
      \label{table1}}
    
\tablehead{
    \colhead{No.}	&
    \colhead{$\alpha$} & 
    \colhead{$\delta$} &
    \colhead{Flux$^1$}  &
     \colhead{Comments}
       \\
    \colhead{}	&
    \colhead{J2000} & 
    \colhead{J2000} & 
    \colhead{(mJy)}  &
    \colhead{} 
   
    }
      
\startdata

1	&  05 35 11.32   	&  $-$05 24 26.4   	& 6.0$\pm$0.6	&  114-426 \\
2	&  05 35 11.32		&  $-$05 24 38.2   	& 88$\pm$10   	&  2MASS 05351131-0524381$^2$ \\  
       %M_H ~ 16.4; M_K ~ 12.8;  M_L ~ 10.1     Muench ( 2002) =  Robberto(2010) # 3880
3	&  05 35 11.66		&  $-$05 24 21.5	& 3.7$\pm$0.6	& OW94 117-421;  2MASS 05351165-0524213	\\
       %M_H ~ 11.8; M_K ~ 11.3;  M_L ~ 10.3      Muench (2002) = Robberto(2010) # 3388
       % Ricci et al. 2008
4	& 05 35 10.53 		& $-$05 24 16.8	& $<$4 		& OW94 106-417;  HH 530; V2202 Ori  \\
	%M_H ~ 12.8; M_K ~ 11.6;  M_L ~ 9.9       Robberto(2010) #1202
5	& 05 35 12.11 		& $-$05 24 33.9	& $<$9 		& OW94 121-434; V2224 Ori \\
	%M_H ~ 13.7; M_K ~ 12.7;  M_L ~ 11.1 
\enddata

\tablecomments{
 {\bf[1]}:   Fluxes are measured on the ALMA dust continuum image.   
 {\bf[2]}:  Source 2 corresponds to 
 ROBb 18 (see Table~3 in  Mann et a. 2014).  \citet{Mann2014} measured a flux of 93.79 mJy
 for this source.  The reprocessed data used in this paper gives a slightly lower flux.   The
 discrepancy is probably due to the fact that the star is at the edge of the primary beam.
               }
\end{deluxetable}

\clearpage

%xxxxxxxx d114-426  2 Oct 2014 xxxxx

\begin{deluxetable}{lccrrrr}
 %\scriptsize{10pt}
 % \tabletypesize{\scriptsize}
 %\setlength{\tabcolsep}{0.02in}
  %\tablewidth{0pt}
  
\centering

\tablecaption{ 
      \CO\  absorption radial velocities and locations in the  d114-426 disk 
      \label{table1}}
    
\tablehead{
    \colhead{V$_{LSR}$}	&
    \colhead{$\alpha$} & 
    \colhead{$\delta$} & 
    \colhead{I(CO)$^a$} &
    \colhead{$\Delta$T(CO)$^a$} &
    \colhead{T$_{bck}^b$}  &
    \colhead{${\Delta T(CO)} \over  T_{bck}$} 
    
       \\
    \colhead{(km s$^{-1}$)} &
    \colhead{J2000} & 
    \colhead{J2000} & 
    \colhead{(Jy)} &
    \colhead{(K)} &
    \colhead{(K)}  &
    \colhead{}
    }
      
\startdata

5.43		&  05 35 11.332	& $-$05 24 26.10	& $-$0.07 & $-$3.0 	& 51 & 0.06 \\
5.86		&  05 35 11.328   	& $-$05 24 26.04   	& $-$0.14 & $-$6.0 	& 59 & 0.10 \\
6.28 		&  05 35 11.336   	& $-$05 24 26.04   	& $-$0.43 & $-$18.4 & 67 & 0.27 \\
6.70		&  05 35 11.336   	& $-$05 24 26.04   	& $-$0.61 & $-$26.1 & 73 & 0.36 \\
7.13		&  05 35 11.334   	& $-$05 24 26.09   	& $-$0.63 & $-$27.0 & 78 & 0.35 \\
7.55		&  05 35 11.331   	& $-$05 24 26.20   	& $-$0.42 & $-$18.0 & 82 & 0.22 \\
7.97		&  05 35 11.326   	& $-$05 24 26.31   	& $-$0.24 & $-$10.3	& 82 & 0.13 \\
8.40		&  05 35 11.322   	& $-$05 24 26.38   	& $-$0.14 & $-$6.0 	& 81 & 0.07 \\
8.82		&  05 35 11.314   	& $-$05 24 26.59   	& $-$0.18 & $-$7.7 	& 77 & 0.10 \\
9.24		&  05 35 11.310   	& $-$05 24 26.72   	& $-$0.28 & $-$12.0 & 71 & 0.17 \\
9.67		&  05 35 11.306   	& $-$05 24 26.81   	& $-$0.26 & $-$11.1 & 63 & 0.17 \\
10.09	&  05 35 11.306   	& $-$05 24 26.81   	& $-$0.15 & $-$6.4 	& 51 & 0.13\\
10.51	&  05 35 11.306   	& $-$05 24 26.81   	& $-$0.05 & $-$2.1 	& 39 & 0.05 \\

\enddata

\tablecomments{
 {\bf[a]}:  I(CO) is the ALMA-DETECTED J=3$-$2 absorption in Jankys;  
 T(CO) is the absorption in brightness temperature units.  
 {\bf[b]}:  The background is estimated by interpolation  of the \CO\ emission in 
 the J=1$-$0 emission from \citet{Shimajiri2011} and the 
 J=2$-$1 emission from \citet{Berne2014}.
               }
\end{deluxetable}

\clearpage

\end{document}